\documentclass[10pt,twocolumn,nofootinbib,nobibnotes,amsmath,amssymb,aps,prd,superscriptaddress,floatfix]{revtex4-2}

\usepackage{amsmath,amssymb,mathrsfs}
\usepackage[dvipsnames]{xcolor}
\usepackage{hyperref}

\hypersetup{
  colorlinks=true,
  linkcolor=BrickRed,
  citecolor=OliveGreen,
  urlcolor=OliveGreen
}

\newcommand{\Z}{\mathbb Z}
\newcommand{\Ztwo}{\mathbb Z_2}

\begin{document}

\title{Satisfying Freed--Witten quantization on smooth
divisors in Calabi--Yau orientifolds}

\author{Mudit Jain}
\email{mudit.jain@rice.edu}
\affiliation{Department of Physics and Astronomy, Rice University, Houston, Texas 77005, U.S.A.}

\date{\today}

\begin{abstract}
Freed--Witten quantization conditions on several divisors wrapped by D-branes must be satisfied with one bulk $B$-field. On a smooth Calabi--Yau threefold $X$ with a holomorphic O3/O7 involution $\sigma$, we show that the common background $B_{2,+}^{\star}=\tfrac12[R]$, where $R$ is the total O7 divisor, permits Freed--Witten period quantization on the full two-cycle lattice of \emph{every smooth connected divisor}, modulo torsion. The worldvolume fluxes are chosen individually and obey the orientifold relation between each divisor and its image. The central step concerns setwise invariant divisors $D$: we prove that the bulk Freed--Witten class $\eta_D^\star=\tfrac12([R]+[D])|_D$ has integer periods on every $\sigma$-even integral two-cycle class, including inherited and autochthonous classes. Its periods on purely $\sigma$-odd classes vanish, so any remaining non-integer periods occur only on classes that are neither $\sigma$-even nor $\sigma$-odd. We show how to choose $\sigma$-odd worldvolume fluxes on these invariant divisors that make all periods integral. For exchanged divisor pairs, quantizing the flux on one member automatically quantizes its orientifold image. On invariant divisors with $h^{2,0}_-(D)=0$, the constructed flux is of type $(1,1)$ and admits a primitive representative; this includes every smooth rigid invariant divisor. We also study zero worldvolume flux on invariant divisors. For $h^{1,1}_-(X)\leq1$, we construct common bulk backgrounds satisfying every inherited intersection period on all such divisors. A binary compatibility test determines whether a common shift exists for larger $h^{1,1}_-(X)$. Non-inherited period tests then determine whether the resulting backgrounds allow zero flux on the full worldvolume lattices.
\end{abstract}

\maketitle


\section{Introduction}
Freed--Witten anomaly cancellation is a necessary condition for a consistent D-brane configuration~\cite{Freed:1999vc}. In Type IIB compactifications on a Calabi--Yau threefold $X$ with a holomorphic O3/O7 involution $\sigma$, the associated worldvolume flux quantization conditions apply both to spacetime-filling D7-branes and to Euclidean D3-branes wrapping divisors.

Our main motivation comes from Euclidean D3-brane instantons, which can generate nonperturbative potentials for K\"ahler moduli and their axionic partners. Whether a given divisor contributes depends on several conditions: its worldvolume flux must be consistently quantized, the orientifold projection must leave the required universal fermion zero modes, and additional neutral or charged modes must not obstruct the amplitude~\cite{Witten:1996bn,Blumenhagen:2009qh}. These conditions are logically distinct. In this paper we address the Freed--Witten period quantization condition. Choices of bulk $B$-field and consistently quantized worldvolume fluxes have been studied in fluxed-instanton and D7-brane constructions~\cite{Collinucci:2010gz,Grimm:2011dj,Cicoli:2011qg,Kerstan:2012cy}. Worldvolume flux can also lift deformation zero modes, allowing non-rigid divisors to support nonperturbative superpotential contributions~\cite{Bianchi:2011qh,Bianchi:2012pn}.

The bulk $B$-field enters the Freed--Witten quantization
condition together with the canonical class of the wrapped
divisor. A choice that compensates the half-integral shift
on one divisor need not compensate it on another.
All divisors share the same bulk background, while their
worldvolume fluxes can differ. These fluxes must also obey
the orientifold relation between each divisor and its image.
On a setwise invariant divisor, this requires the
gauge-invariant flux $\mathcal F_D$ to be
$\sigma$-odd~\cite{Grimm:2011dj}. We therefore ask whether
one bulk $B$-field permits consistently quantized fluxes
on every smooth connected divisor, together with its
orientifold image.

The background $B_{2,+}^{\star}=\tfrac12[R]$, where $R$ is the total O7 divisor, has been used to permit vanishing gauge-invariant flux on D7 stacks supported on the O7 locus~\cite{Demirtas:2021nlu,McAllister:2024lnt}. Other common backgrounds can also work in particular geometries~\cite{Louis:2012nb,Crino:2022zjk}. Here we establish a general guarantee: under the assumptions stated in Sec.~\ref{sec:quantization}, the single background $B_{2,+}^{\star}$ permits Freed--Witten period quantization on the full two-cycle lattice of every smooth connected divisor. The worldvolume fluxes are chosen individually and obey the orientifold image relation. No restriction on $h^{1,1}_-(X)$ is required.

The essential step concerns smooth connected setwise invariant divisors $D\subset X$, with $\sigma(D)=D$. Such a divisor need not be supported on the O7 locus. We prove that the corresponding bulk Freed--Witten class $\eta_D^\star=\tfrac12([R]+[D])|_D$ (see Eq.~\eqref{eq:eta}) has integer periods on every invariant ($\sigma$-even) integral worldvolume two-cycle class. This includes inherited and autochthonous classes. The relevant condition is invariance of the class; its representative need not be invariant or holomorphic. The proof combines the canonical-class form of Wu's formula with a classical intersection identity for smooth involutions~\cite{GompfStipsicz1999,Arnold:1971ovals,Degtyarev:1997pv}.

Since $\eta_D^\star$ is $\sigma$-even, its periods on purely $\sigma$-odd classes vanish. Any non-integer periods can therefore occur only on integral classes of mixed $\sigma$-parity. We show how to repair these periods with a $\sigma$-odd worldvolume flux. The construction uses an integral class $\ell_D$ whose $\sigma$-even projection is $\eta_D^\star$. Taking its $\sigma$-odd projection as $\mathcal F_D$ makes $\eta_D^\star+\mathcal F_D=\ell_D$ integral on the full worldvolume lattice. If $h^{2,0}_-(D)=0$, the flux is of type $(1,1)$ and admits a primitive representative for an invariant K\"ahler metric. This includes every smooth rigid invariant divisor.

For an exchanged pair $D'=\sigma(D)\neq D$, the construction is simpler. One chooses a quantized flux on $D$ and assigns $\mathcal F_{D'}=-\sigma^*\mathcal F_D$. At $B_{2,+}^{\star}$, the image flux then satisfies quantization automatically. Together with the invariant-divisor construction, this establishes the common-background result for any collection of smooth divisors and their images. Primitivity on exchanged divisors must, however, be checked separately.

We also study when setwise invariant divisors permit the stronger choice $\mathcal F_D=0$. This choice is particularly relevant in statistical studies of axions, where contributions from many instanton divisors are included at once~\cite{Demirtas:2018akl,Mehta:2021pwf,Sheridan:2024vtt,Jain:2025vfh}. We give an exact test on arbitrary integral worldvolume classes. When $h^{1,1}_-(X)=0$, every inherited intersection class is invariant, so $B_{2,+}^{\star}$ already satisfies every inherited period on every smooth connected invariant divisor. This requires no assumptions on $h^{2,0}(D)$ or $h^{1,1}_-(D)$. When bulk classes are exchanged, integer periods on their invariant sums need not imply integer periods on the individual classes. We construct common half-sum shifts of $B_{2,+}^{\star}$ that satisfy every inherited period when $h^{1,1}_-(X)\leq1$. For several exchanged pairs, a binary compatibility test determines which common shifts work. Retaining zero flux on the full worldvolume also requires the non-inherited periods to be integral.

Section~\ref{sec:quantization} fixes the flux conventions and cycle notation. Section~\ref{sec:theorem} proves the invariant-period theorem. Section~\ref{sec:wv-flux} constructs the worldvolume fluxes on invariant divisors and exchanged pairs, establishes the common-background result, and discusses the supersymmetry conditions. Sections~\ref{sec:scope} and \ref{sec:corollaries} study zero flux on invariant divisors, first on the full worldvolume lattice and then on inherited lattices. Appendix~\ref{sec:applications} describes the binary search over general invariant backgrounds. The remaining appendices give worked computations and explicit geometric examples.

\section{Flux quantization and worldvolume classes}
\label{sec:quantization}
Let $\sigma$ be a holomorphic involution of a smooth projective Calabi--Yau threefold $X$, with $\sigma^*\Omega_X=-\Omega_X$. Its fixed locus consists of smooth divisor components and isolated points, corresponding to O7- and O3-planes. We write
\begin{equation}
 R=\bigcup_{i=1}^{n_R}R_i,
 \qquad [R]=\sum_{i=1}^{n_R}[R_i],
 \label{eq:R}
\end{equation}
where the $R_i$ are the mutually disjoint O7 components, each counted once. Each $R_i$ is fixed pointwise, so each individual class $[R_i]$ is invariant.

For D7-branes or an ED3 instanton wrapping a smooth connected divisor $D$, let $\iota:D\hookrightarrow X$ denote the inclusion. The pullback $\iota^*$ restricts bulk forms and their cohomology classes to $D$. In the usual flux normalization,
\[
 \mathcal F_D=F_D-\iota^*B_2,
\]
the Freed--Witten quantization condition is
\begin{equation}
 F_D+\tfrac12c_1(K_D)\in H^2(D,\Z).
 \label{eq:FW}
\end{equation}
Here $F_D$ is the worldvolume gauge-field strength, $B_2$ is the bulk Neveu--Schwarz two-form, and $\mathcal F_D$ is their gauge-invariant combination. The canonical line bundle $K_D$ has holomorphic two-forms as its local sections, and $c_1(K_D)$ is the canonical class of type $(1,1)$.

To express the canonical class in terms of bulk divisor classes, we use adjunction. Since $K_X$ is trivial, the canonical class is just the restriction of $[D]$:
\begin{equation}
 K_D\simeq\mathcal O_X(D)|_D,
 \qquad c_1(K_D)=\iota^*[D].
 \label{eq:adjunction}
\end{equation}
In particular, $c_1(K_D)$ is integral. When $D$ is setwise invariant, it is also $\sigma$-even.

Let $D'=\sigma(D)$ be the image divisor. We consider Abelian worldvolume flux classes satisfying the orientifold image relation
\begin{equation}
 \mathcal F_{D'}=-\sigma^*\mathcal F_D.
 \label{eq:image-flux}
\end{equation}
Here $\sigma^*$ transports classes from $D$ to $D'$. For a setwise invariant divisor, this becomes $\sigma^*\mathcal F_D=-\mathcal F_D$, the parity condition relevant for $O(1)$ instantons~\cite{Grimm:2011dj}. We ask whether one common bulk $B$-field permits quantized fluxes satisfying this relation on every smooth connected divisor.

We set the continuous $\sigma$-odd bulk background $B_{2,-}$ to zero and work with closed worldvolume fluxes.\footnote{The underlying topological condition is $W_3(D)+\iota^*[H_3]=0$~\cite{Freed:1999vc}, where $W_3(D)$ is the third integral Stiefel--Whitney class of the tangent bundle of $D$, and $[H_3]$ is the integral cohomology class of the bulk NSNS three-form flux $H_3$. For a complex surface, $W_3(D)=0$, so this requires $\iota^*[H_3]=0$. In the source-free Abelian setting, $d\mathcal F_D=-\iota^*H_3$. Closed worldvolume fluxes therefore make the restricted three-form vanish, so the topological condition holds modulo torsion. The bulk three-form flux itself need not vanish.} The bulk $B$-field and canonical contributions to the Freed--Witten condition combine into
\begin{equation}
 \boxed{\begin{aligned}
  \eta_D&:=\iota^*B_{2,+}+\tfrac12c_1(K_D)\\
        &=\iota^*\bigl(B_{2,+}+\tfrac12[D]\bigr).
 \end{aligned}}
 \label{eq:eta}
\end{equation}
We refer to $\eta_D$ as the \emph{bulk Freed--Witten class} on $D$, since both contributions are restrictions of bulk classes. The full quantization condition is
\begin{equation}
 \eta_D+\mathcal F_D\in H^2(D,\Z).
 \label{eq:FW-shifted}
\end{equation}
We first consider setwise invariant divisors. For the discrete backgrounds considered here, both terms in $\eta_D$ are $\sigma$-even and of type $(1,1)$. However, the $\sigma$-even and $\sigma$-odd projections of an integral class need not themselves be integral. Thus $\eta_D+\mathcal F_D$ can be integral even when its $\sigma$-even part $\eta_D$ is not. The $\sigma$-odd part of the integral sum is then $\mathcal F_D$. The choice $\mathcal F_D=0$ instead requires
\[
    \eta_D\in H^{1,1}_{+}(D,\Z)\,.
\]
Being of type $(1,1)$ and $\sigma$-even does not by itself establish integrality. Nor does the Hodge type restrict the quantization test to holomorphic curve-cycle classes. The condition must hold on every integral two-cycle, as described below. 

We test quantization through periods on integral worldvolume two-cycle classes $\gamma\in H_2(D,\Z)$. For a closed two-form $\alpha$, we use $\int_\gamma\alpha$ to denote its integral over any representative of $\gamma$. The result depends only on the class, and we use the same notation when $\alpha$ denotes its cohomology class. The required condition is an integer period of $\eta_D+\mathcal F_D$ on \emph{every} integral worldvolume two-cycle in the covering Calabi--Yau, including classes without holomorphic curve-cycle representatives.\footnote{Throughout, homology and integral cohomology lattices are understood modulo torsion, and integrality means integer periods on these free lattices.} 

The action of $\sigma$ on the worldvolume classes of $D$ determines which periods are guaranteed. An invariant ($\sigma$-even) class satisfies $\sigma_*\gamma=\gamma$. This is a condition on the class and does not require an invariant or holomorphic representative. A $\sigma$-odd class satisfies $\sigma_*\gamma=-\gamma$ and has zero period of the invariant class $\eta_D$. We say that an integral class has \emph{mixed $\sigma$-parity} when it is neither $\sigma$-even nor $\sigma$-odd.

A $\sigma$-odd flux has zero period on every $\sigma$-even class. It therefore cannot repair a half-integer period of $\eta_D$ on such a class. Integer periods on all $\sigma$-even integral classes are necessary for a $\sigma$-odd flux to satisfy Eq.~\eqref{eq:FW-shifted}. We will show that this necessary condition is also sufficient.


\section{The common background}
\label{sec:theorem}
The discrete choice
\begin{equation}
 \boxed{B_{2,+}^{\star}=\tfrac12[R]}
 \label{eq:Bstar}
\end{equation}
is invariant under the geometric involution, since $\sigma^*[R]=[R]$. The full orientifold also includes worldsheet orientation reversal, which sends $B_2\mapsto-\sigma^*B_2$. Backgrounds differing by integral shifts are gauge equivalent, so a geometrically invariant background is allowed when $2B_{2,+}$ is integral. This holds here because $2B_{2,+}^{\star}=[R]$. In what follows, $\eta_D^\star$ denotes the bulk Freed--Witten class evaluated at this background.

\begin{samepage}

\medskip
\noindent\textbf{Theorem 1.}
\emph{Let $D\subset X$ be a smooth connected divisor satisfying $\sigma(D)=D$. For every invariant integral worldvolume two-cycle class $\gamma\in H_2(D,\Z)$,}
\begin{equation}
 \boxed{
 \int_\gamma\eta_D^{\star}\in\Z.
 }
 \label{eq:invariant-parity}
\end{equation}
\emph{Thus the common background $B_{2,+}^{\star}=\tfrac12[R]$ gives integer Freed--Witten periods on every such class. No invariant or holomorphic representative is required. Purely $\sigma$-odd integral classes have zero period. If $D$ is an O7 component, the condition holds on every integral worldvolume two-cycle.}
\end{samepage}

\subsection{O7 components}
\label{subsec:on-O7}

If $D$ is an O7 component, write $R=D+R'$. The other fixed components do not meet $D$, so $\iota^*[R']=0$. Adjunction gives
\begin{equation}
 \eta_D^{\star}
 =\tfrac12\iota^*[R]+\tfrac12c_1(K_D)
 =\iota^*[D].
 \label{eq:O7-integral}
\end{equation}
This is an integral class and has integer periods on every integral two-cycle. It explains the familiar cancellation on D7 stacks supported on O7 components~\cite{Demirtas:2021nlu,McAllister:2024lnt}.

\subsection{Invariant divisors that are not O7 components}
\label{subsec:not-O7}

Suppose $D$ is not an O7 component. It is preserved setwise, but is not fixed pointwise, so its induced involution is nontrivial. We first relate its fixed curves to the O7 locus, and then use two intersection identities to prove the required integrality.

\subsubsection{Fixed curves on the divisor}

Let $D_r$ be the union of the fixed curves on $D$, each counted once. Near an O7 component, choose local coordinates with
\begin{equation}
 \sigma(z_1,z_2,z_3)=(-z_1,z_2,z_3),
 \qquad R=\{z_1=0\}.
\end{equation}
Setwise invariance allows a local defining equation for $D$ to be chosen either even or odd in $z_1$. A defining equation odd in $z_1$ would be divisible by $z_1$, and smoothness would then make $D$ coincide locally with the O7 component. Since $D$ is not an O7 component, we may therefore use a defining equation even in $z_1$. Moreover, smoothness requires a nonzero derivative in a direction tangent to $R$. After relabeling $z_2$ and $z_3$ if necessary, the local defining equation for $D$ can therefore be written as $z_2=f(z_1^2,z_3)$. Here $(z_1,z_3)$ are local coordinates on $D$, so $D$ meets $R$ transversely and the fixed curve is cut out once by $z_1=0$. Consequently,
\begin{equation}
 [D_r]=\iota^*[R]\in H^2(D,\Z).
 \label{eq:D_r}
\end{equation}
At an O3 point the induced action is locally $(u,v)\mapsto(-u,-v)$. This gives an isolated fixed point, with no curve contribution.\footnote{The same fixed-curve class enters the ramification formula for the surface quotient $p:D\to S=D/\sigma$: $c_1(K_D)=p^*c_1(K_S)+[D_r]$. See Ref.~\cite{Calabri2005}. The quotient has ordinary double points at images of isolated fixed points, where its canonical bundle remains defined. These points introduce no additional curve term.}

\subsubsection{Two intersection identities and the proof}

Write $\gamma^2$ for the self-intersection number of $\gamma$ and $\gamma\cdot\sigma_*\gamma$ for its intersection number with its image. These numbers are computed using smooth oriented surface representatives in $D$, perturbed without changing their classes so that they meet transversely at isolated points. Each intersection is counted with its orientation sign. For $\gamma^2$, we use two representatives of the same class. These perturbations need not preserve holomorphicity, even when a chosen representative is a rigid holomorphic curve.

The canonical-class form of Wu's formula is~\cite{GompfStipsicz1999}
\begin{equation}
 \int_\gamma c_1(K_D)\equiv\gamma^2\pmod2.
 \label{eq:Wu}
\end{equation}
For a smooth holomorphic curve $C$, the same congruence follows from adjunction:
\[
 \int_Cc_1(K_D)
 =2g(C)-2-C^2
 \equiv C^2\pmod2.
\]
Wu's formula establishes this congruence for every integral two-cycle class, without requiring a holomorphic representative.

The second identity is~\cite{Arnold:1971ovals,Degtyarev:1997pv}
\begin{equation}
 \int_\gamma[D_r]
 \equiv\gamma\cdot\sigma_*\gamma\pmod2.
 \label{eq:Arnold}
\end{equation}
This is the four-dimensional case of Arnold's fixed-locus intersection identity. See Proposition~3.2.3 of Ref.~\cite{Degtyarev:1997pv}. Intersections of a surface with its image away from the fixed locus occur in pairs. The unpaired contribution is its intersection with the fixed curves. Surface representatives can avoid the isolated fixed points.

Adding Eqs.~\eqref{eq:Wu} and \eqref{eq:Arnold}, and using Eq.~\eqref{eq:D_r}, gives
\begin{equation}
 2\int_\gamma\eta_D^{\star}
 \equiv\gamma^2+\gamma\cdot\sigma_*\gamma\pmod2.
 \label{eq:general-parity}
\end{equation}
For an invariant class, the right-hand side is $2\gamma^2$, which is divisible by two. This proves Theorem~1. Invariance is used only at this last step. No invariant representative was assumed. Equation~\eqref{eq:general-parity} also holds on O7 components. There $\sigma_*\gamma=\gamma$, so its right-hand side is even, while Eq.~\eqref{eq:O7-integral} makes its left-hand side even as well.

For involutions with only O3 planes, there are no O7 components and thus no fixed curves on $D$. Equation~\eqref{eq:Arnold} then gives $\gamma\cdot\sigma_*\gamma\equiv0\pmod2$. An invariant class therefore has self-intersection divisible by two. Equation~\eqref{eq:Wu} then makes its canonical period divisible by two as well. The theorem holds with $B_{2,+}^{\star}=0$.

An immediate consequence gives a sufficient condition for zero flux on the full worldvolume lattice.

\medskip
\noindent\textbf{Corollary 1 (zero odd second cohomology).}
\emph{If a smooth connected invariant divisor $D$ satisfies}
\begin{equation}
 \boxed{h^{1,1}_{-}(D)=h^{2,0}_{-}(D)=0,}
 \label{eq:full-Hodge-condition}
\end{equation}
\emph{then $B_{2,+}^{\star}$ satisfies every integral worldvolume period condition with $\mathcal F_D=0$.}

Indeed, $h^{0,2}_{-}(D)=0$ as well, so the $\sigma$-odd part of $H^2(D,\mathbb C)$ vanishes. Every class in the free integral two-cycle lattice is therefore invariant, and Theorem~1 applies to the entire lattice. No vanishing of $h^{2,0}_{+}(D)$ is needed. For a rigid divisor, the condition reduces to $h^{1,1}_{-}(D)=0$.

\section{Worldvolume fluxes and full quantization}
\label{sec:wv-flux}
For a setwise invariant divisor, we first show that integer periods on invariant classes are necessary and sufficient for completing quantization with a $\sigma$-odd flux. Theorem~1 guarantees these periods at $B_{2,+}^\star$. We then treat exchanged pairs and combine the two constructions.

\subsection{Constructing the worldvolume fluxes}

\medskip
\noindent\textbf{Proposition 1 (odd-flux completion).}
\emph{Let $D$ be setwise invariant and let $\eta\in H^2(D,\mathbb Q)$ be invariant. There exists a $\sigma$-odd rational class $\mathcal F$ for which $\eta+\mathcal F$ has integer periods on every integral two-cycle if and only if $\eta$ has integer periods on every invariant integral two-cycle class.}

\medskip
Necessity follows because a $\sigma$-odd class has zero period on an invariant cycle. To prove sufficiency, write
\begin{equation}
 \begin{aligned}
 L&=H_2(D,\Z)/\operatorname{Tor}H_2(D,\Z),\\
 L_+&=\{\gamma\in L:\sigma_*\gamma=\gamma\}.
 \end{aligned}
 \label{eq:worldvolume-invariant-lattice}
\end{equation}
The quotient removes torsion, so $L$ has a finite integral basis. If a nonzero integer multiple $n\gamma$ lies in $L_+$, then $n(\sigma_*\gamma-\gamma)=0$. Since $L$ is torsion-free, $\gamma$ itself lies in $L_+$. This property ensures that an integral basis of $L_+$ can be extended to a full integral basis of $L$. Choose such a basis,
\begin{equation}
 \gamma_1,\ldots,\gamma_k,\mu_1,\ldots,\mu_s,
\end{equation}
where the $\gamma_a$ form a basis of $L_+$. The added classes $\mu_b$ need not have definite $\sigma$-parity.

An integral cohomology class can be specified by assigning integer periods on a full integral two-cycle basis. Choose $\ell$ to agree with $\eta$ on the invariant basis cycles and to have zero periods on the added cycles:
\begin{equation}
 \int_{\gamma_a}\ell=\int_{\gamma_a}\eta\in\Z,
 \qquad
 \int_{\mu_b}\ell=0.
 \label{eq:integral-period-extension}
\end{equation}
For any $\gamma\in L$, the sum $\gamma+\sigma_*\gamma$ is invariant. Consequently,
\begin{align}
 \int_\gamma(\ell+\sigma^*\ell)
 &=\int_{\gamma+\sigma_*\gamma}\ell\nonumber\\
 &=\int_{\gamma+\sigma_*\gamma}\eta
 =2\int_\gamma\eta.
\end{align}
Thus $\tfrac12(\ell+\sigma^*\ell)=\eta$ in rational cohomology: $\eta$ is the $\sigma$-even projection of $\ell$. Define the flux to be its $\sigma$-odd projection:
\begin{equation}
 \mathcal F=\tfrac12(\ell-\sigma^*\ell),
 \qquad
 \eta+\mathcal F=\ell.
 \label{eq:odd-flux-lift}
\end{equation}
The flux is $\sigma$-odd and the sum is integral, proving the proposition. The zero periods assigned to the $\mu_b$ were arbitrary; any integer choices give a solution.

\medskip
\noindent\textbf{Theorem 2 (a common background for all
smooth divisors).}
\emph{Under the assumptions of Sec.~\ref{sec:quantization},
consider any collection of smooth connected divisors
together with their orientifold images. The single bulk
background $B_{2,+}^{\star}=\tfrac12[R]$ admits
worldvolume flux classes satisfying}
\begin{equation}
 \boxed{\begin{aligned}
  \mathcal F_{\sigma(D)}&=-\sigma^*\mathcal F_D,\\
  \eta_D^\star+\mathcal F_D
    &\in H^2(D,\Z)_{\mathrm{free}}
 \end{aligned}}
 \label{eq:common-odd-flux}
\end{equation}
\emph{on every divisor in the collection. The fluxes
can be chosen with $2\mathcal F_D$ integral. Quantization
holds on the full worldvolume two-cycle lattices,
including inherited and autochthonous classes, with
no restriction on $h^{1,1}_-(X)$.}

\medskip
If $D$ is setwise invariant, Theorem~1 supplies the integer invariant-cycle periods required by Proposition~1. The proposition produces an integral class $\ell_D$ whose $\sigma$-even projection is $\eta_D^\star$. Its $\sigma$-odd projection $\mathcal F_D=\ell_D-\eta_D^\star$ satisfies both conditions.

Now suppose $D'=\sigma(D)\neq D$. Choose any integral class $\ell_D$ on one member of the pair and set
\begin{equation}
 \mathcal F_D=\ell_D-\eta_D^\star,
 \qquad
 \mathcal F_{D'}=-\sigma^*\mathcal F_D.
 \label{eq:pair-flux-construction}
\end{equation}
On $D$, quantization holds because $\eta_D^\star+\mathcal F_D=\ell_D$ is integral by construction. It remains to check the image divisor. Since $\eta_{D'}^\star=\sigma^*\eta_D^\star$, its image satisfies
\begin{align}
 \eta_{D'}^\star+\mathcal F_{D'}
 &=\sigma^*\bigl(\eta_D^\star-\mathcal F_D\bigr)\nonumber\\
 &=\sigma^*\bigl(2\eta_D^\star-\ell_D\bigr).
\end{align}
This is integral because
\[
 2\eta_D^\star=\iota^*\bigl([R]+[D]\bigr)
\]
is integral.

The exchanged-pair construction works for any discrete $\sigma$-even background with integral $2B_{2,+}$. Theorem~1 ensures that $B_{2,+}^\star$ also permits the required odd fluxes on invariant divisors. The two constructions therefore provide one common background for the entire collection. In both cases, $2\mathcal F_D$ is integral. On O7 components and on the invariant divisors of Corollary~1, $\mathcal F_D=0$ already satisfies quantization.

On a setwise invariant divisor, starting from any one solution $\mathcal F_D^{(0)}$, we obtain the full family of quantized $\sigma$-odd fluxes by adding an arbitrary integral $\sigma$-odd class:
\begin{equation}
 \mathcal F_D=\mathcal F_D^{(0)}+\xi_D,
 \qquad
 \xi_D\in H^2_-(D,\Z)_{\mathrm{free}}.
 \label{eq:odd-flux-family}
\end{equation}
The construction does not require these fluxes to be inherited from the bulk. Implementing it can require the full integral worldvolume lattice and its involution action, beyond the bulk toric intersection data. Appendix~\ref{app:odd-flux-construction} gives the corresponding integer system and a simple example.

\subsection{When the flux satisfies the supersymmetry conditions}

We first consider the $\sigma$-odd flux on a setwise invariant divisor. For a supersymmetric instanton, this flux must also be of type $(1,1)$ and primitive~\cite{Grimm:2011dj}. A sufficient condition is the absence of $\sigma$-odd holomorphic two-forms.

\medskip
\noindent\textbf{Corollary 2 (odd $(1,1)$ flux).}
\emph{On a smooth connected setwise invariant divisor $D$ with $h^{2,0}_-(D)=0$, the flux in Theorem~2 is of type $(1,1)$ and admits a primitive representative for an invariant K\"ahler form. In particular, this holds on every smooth rigid invariant divisor, without requiring $h^{1,1}_-(D)=0$.}

\medskip
The flux class $\mathcal F_D$ is real and $\sigma$-odd. For a $\sigma$-even K\"ahler form $J$, parity gives
\begin{equation}
 \int_D\iota^*J\wedge\mathcal F_D=0.
 \label{eq:odd-flux-primitivity}
\end{equation}
On a connected K\"ahler surface, the component of a harmonic two-form proportional to the K\"ahler form has a constant coefficient. Equation~\eqref{eq:odd-flux-primitivity} forces this coefficient to vanish, so the harmonic representative of $\mathcal F_D$ is primitive. If, in addition, $h^{2,0}_-(D)=0$, its $(2,0)$ and $(0,2)$ components vanish, so this representative is also of type $(1,1)$. When $h^{2,0}_-(D)\neq0$, the $(1,1)$ condition must be tested separately.

For an exchanged pair, choosing $\ell_D=0$ in Eq.~\eqref{eq:pair-flux-construction} gives $\mathcal F_D=-\eta_D^\star$ and $\mathcal F_{D'}=\eta_{D'}^\star$. These classes are of type $(1,1)$. The image relation does not, however, force $\int_D\iota^*J\wedge\mathcal F_D=0$ on either divisor individually. Primitivity must therefore be checked separately.

\section{Zero flux on invariant divisors}
\label{sec:scope}
We now return to setwise invariant divisors and ask when $\eta_D^\star$ itself has integer periods on every integral two-cycle, allowing $\mathcal F_D=0$. Corollary~1 covers divisors with no $\sigma$-odd second cohomology. On any other invariant divisor, Theorem~1 still guarantees the $\sigma$-even periods, including invariant autochthonous classes, and the $\sigma$-odd periods vanish. The remaining tests are on integral classes of mixed $\sigma$-parity.

For an arbitrary integral worldvolume class $\gamma$, define $\delta_\gamma=\gamma-\sigma_*\gamma$. Preservation of the intersection pairing gives
\begin{equation}
 \delta_\gamma^2
 =2\bigl(\gamma^2-\gamma\cdot\sigma_*\gamma\bigr).
 \label{eq:odd-square}
\end{equation}
Equation~\eqref{eq:general-parity} therefore becomes
\begin{equation}
 2\int_\gamma\eta_D^{\star}
 \equiv\tfrac12\delta_\gamma^2\pmod2.
 \label{eq:difference-parity}
\end{equation}
Therefore, the exact period test is
\begin{equation}
 \boxed{\int_\gamma\eta_D^{\star}\in\Z
 \quad\Longleftrightarrow\quad
 \delta_\gamma^2\in4\Z.}
 \label{eq:full-criterion}
\end{equation}
An integral class of mixed $\sigma$-parity need not violate quantization: Eq.~\eqref{eq:full-criterion} determines whether its period is integral.

If curves $C$ and $\sigma(C)$ are exchanged, invariance of $\eta_D^{\star}$ gives
\begin{equation}
 \int_C\eta_D^{\star}
 =\int_{\sigma(C)}\eta_D^{\star}
 =\tfrac12\int_{C+\sigma(C)}\eta_D^{\star}.
 \label{eq:pair-period}
\end{equation}
An integer period on the sum can coexist with half-integer periods on the components. Both component periods must satisfy quantization.

Appendix~\ref{app:double-octic} gives an explicit example in which every inherited period is integral, while some autochthonous periods are half-integer.

\section{Zero flux on inherited lattices}
\label{sec:corollaries}
We continue to impose $\mathcal F_D=0$ and first test the intersection classes inherited from the bulk. Theorem~1 guarantees integer periods on invariant inherited classes. We identify the individual tests for exchanged classes and show when and how a common bulk shift can make every inherited period integral on all setwise invariant target divisors. Zero flux on the full worldvolume additionally requires the autochthonous periods to be integral.

\subsection{The inherited corollary}
\label{subsec:inherited-corollary}

For an arbitrary integral bulk divisor class $[E]\in H^2(X,\Z)$, invariant or otherwise, let $\gamma_E$ denote its inherited intersection class on $D$. It is characterized, for closed two-forms $\alpha$, by
\begin{equation}
 \int_{\gamma_E}\alpha
 =\int_D\iota^*[E]\wedge\alpha.
 \label{eq:inherited-class}
\end{equation}
If the geometric intersection $D\cap E$ consists of curves $C_j$ with multiplicities $n_j$, then $\gamma_E=\sum_j n_j[C_j]$ is their homology class. Integer combinations of the classes $\gamma_E$, as $[E]$ ranges over the integral bulk divisor lattice, form the \emph{inherited intersection curve lattice} of $D$. An individual component $C_j$ need not be inherited. Classes outside this lattice are non-inherited, or autochthonous.

This definition also includes $E=D$: by adjunction, $\gamma_D$ corresponds to $\iota^*[D]=c_1(K_D)$. It is defined by restriction even if $[D]$ has no other holomorphic representative with which to intersect $D$.

At $B_{2,+}^{\star}$, define the inherited period
\begin{align}
 I(D,E)&:=\int_{\gamma_E}\eta_D^{\star}\nonumber\\
 &=\tfrac12\int_X[D]\,[E]\,\bigl([D]+[R]\bigr).
 \label{eq:I}
\end{align}
This period may be half-integer, whereas Freed--Witten quantization requires it to be an integer.

\medskip
\noindent\textbf{Corollary 3.}
\emph{If the inherited class $\gamma_E$ is invariant, then}
\begin{equation}
 I(D,E)\in\Z.
 \label{eq:theorem}
\end{equation}
\emph{In particular, this holds for every invariant integral bulk class $[E]$, including $E=D$. When $h^{1,1}_{-}(X)=0$, it holds on the full inherited intersection curve lattice of $D$.}

Indeed, restriction to invariant $D$ commutes with the involution. An invariant bulk class therefore restricts to an invariant worldvolume class. A non-invariant bulk class can also restrict to an invariant class if its $\sigma$-odd projection restricts to zero. No condition on $h^{2,0}(D)$ or $h^{1,1}_{-}(D)$ is needed for this corollary. A purely $\sigma$-odd bulk class has zero period trivially.

\subsection{A bulk shift for an exchanged pair}
\label{subsec:exchanged-periods}

The target divisor $D$ remains setwise invariant. Let $[e]\leftrightarrow[e']$ be an exchanged pair of integral bulk divisor classes. Their inherited classes $\gamma_e$ and $\gamma_{e'}=\sigma_*\gamma_e$ have equal periods of $\eta_D^\star$, but these periods may be half-integer.

The two intersection numbers entering Eq.~\eqref{eq:general-parity} are
\[
 \gamma_e^2=\int_X[D][e]^2,
 \qquad
 \gamma_e\cdot\sigma_*\gamma_e
 =\int_X[D][e][e'].
\]
That equation therefore reads
\begin{equation}
 2\int_{\gamma_e}\eta_D^\star
 = 2I(D,e) \equiv
 \int_X[D][e]\bigl([e]+[e']\bigr)\pmod2.
 \label{eq:pair-shift-parity}
\end{equation}
In other words, the periods of $\eta_D^\star$ and $\tfrac12\iota^*([e]+[e'])$ on $\gamma_e$ differ by an integer. Subtracting this half-sum from the bulk background therefore makes the inherited period integral:
\begin{equation}
 \tilde B_{2,+}^{\star}
 =\tfrac12[R]-\tfrac12\bigl([e]+[e']\bigr).
 \label{eq:one-exchange-background}
\end{equation}
Since this background is $\sigma$-even, the image class $\sigma_*\gamma_e$ has the same period and passes the same test.

The periods on invariant integral worldvolume classes remain integers, since
\begin{equation}
 \tfrac12\int_\gamma\iota^*\bigl([e]+[e']\bigr)
 =\int_\gamma\iota^*[e]\in\Z
 \qquad(\sigma_*\gamma=\gamma).
 \label{eq:preserve-invariant}
\end{equation}
On an O7 component every worldvolume class is invariant, so integrality is preserved on its full lattice. Purely $\sigma$-odd periods remain zero. The background $\tilde B_{2,+}^{\star}$ also obeys the orientifold condition: it is $\sigma$-even and twice its class is integral.

\medskip
\noindent\textbf{Theorem 3.}
\emph{Suppose $\sigma$ exchanges one pair $[e]\leftrightarrow[e']$ in an integral basis of divisor classes on $X$ and leaves all other basis classes invariant. Then the common background \eqref{eq:one-exchange-background} satisfies Freed--Witten period quantization on the full inherited intersection curve lattice of every smooth connected setwise invariant divisor $D$.}

\medskip
The inherited periods associated with $[e]$ and $[e']$ are integral by the preceding calculation. All remaining basis classes are invariant, and their periods remain integral by Eq.~\eqref{eq:preserve-invariant}. Since every inherited class is an integer combination of these basis intersections, this proves the theorem.

If several pairs of basis classes are exchanged, shifting the background for one pair can change the periods of another. We must therefore choose a common combination of shifts that makes every inherited period integral on all target divisors.

\subsection{Choosing a common combination}
\label{subsec:shifts}

Choose an integral divisor basis in which $\sigma$ exchanges $m$ pairs and every remaining class is $\sigma$-even or $\sigma$-odd.\footnote{Such a basis exists for every integral involution; see Proposition~3.2 and Appendix~A of Ref.~\cite{Borovoi:2021component}.} Let $[A_\alpha]$, $\alpha=1,\ldots,m$, denote one class from each exchanged pair. Fixed basis classes contribute only integral shifts, while $\sigma$-odd basis classes contribute no shift. Only the half-sums of exchanged pairs therefore need to be tested.

For each pair, we may include or omit the shift $\tfrac12([A_\alpha]+\sigma^*[A_\alpha])$. There are therefore $2^m$ possible combinations. Write
\begin{align}
 [A]&=\sum_{\alpha=1}^{m}z^\alpha[A_\alpha],
 \qquad z^\alpha\in\Ztwo,
 \label{eq:shift-coordinates}\\
 \tilde B_{2,+}^{\star}
 &=\tfrac12[R]-\tfrac12\bigl([A]+\sigma^*[A]\bigr).
 \label{eq:corrected-B}
\end{align}
The corresponding bulk Freed--Witten class is
\begin{equation}
 \tilde\eta_D^\star
 =\eta_D^\star-\tfrac12\iota^*\bigl([A]+\sigma^*[A]\bigr).
 \label{eq:corrected-eta}
\end{equation}
For every candidate background, the invariant periods remain integral by Eq.~\eqref{eq:preserve-invariant}. It suffices to test one class from each exchanged pair: its image has the same period because the background $\tilde B_{2,+}^{\star}$ is $\sigma$-even. All remaining basis classes have definite $\sigma$-parity, so their inherited periods are already integral or zero.

For $m=0$, take $[A]=0$. For $m=1$, take $[A]=[A_1]$, using the one-pair argument of Theorem~3. Each exchanged pair contributes one $\sigma$-odd direction, so $m\leq h^{1,1}_-(X)$. Thus one common shift always works when $h^{1,1}_-(X)\leq1$, and also at higher odd rank when at most one pair is exchanged in this basis.

For several independent shifts, their effects on all selected periods must be included. For each target divisor $D_i$, define
\begin{equation}
 \begin{aligned}
 M^{(i)}_{\alpha\beta}
 &:=\int_X[D_i][A_\alpha]
       \bigl([A_\beta]+\sigma^*[A_\beta]\bigr)\bmod2,\\
 d^{(i)}_\alpha
 &:=2I(D_i,A_\alpha)\bmod2
  =M^{(i)}_{\alpha\alpha}.
 \end{aligned}
 \label{eq:M}
\end{equation}
These matrices are symmetric: the term containing $[A_\alpha][A_\beta]$ is manifestly symmetric, and applying $\sigma$ to the other term exchanges $\alpha$ and $\beta$ while leaving $[D_i]$ invariant. The equality in the second line of Eq.~\eqref{eq:M} follows from Eq.~\eqref{eq:pair-shift-parity}. An entry $d^{(i)}_\alpha=1$ marks a half-integer period at $B_{2,+}^\star$. An entry $M^{(i)}_{\alpha\beta}=1$ means that including shift $\beta$ changes period $\alpha$ by a half-integer. All inherited periods on $D_i$, with this background,  are therefore integral precisely when
\begin{equation}
 M^{(i)}z=d^{(i)}\quad\text{over }\Ztwo.
 \label{eq:single-shift-system}
\end{equation}

One common background requires the same $z$ for every target. Stack their equations into the single binary system
\begin{equation}
 \underbrace{
 \begin{pmatrix}M^{(1)}\\M^{(2)}\\\vdots\end{pmatrix}
 }_{\mathcal M}
 z=
 \underbrace{
 \begin{pmatrix}d^{(1)}\\d^{(2)}\\\vdots\end{pmatrix}
 }_{\mathbf d}.
 \label{eq:common-system}
\end{equation}
Gaussian elimination using XOR row additions finds every compatible combination, if any. Appendix~\ref{app:lattice-shifts} illustrates why including all half-sums need not work, even when a smaller combination does. For $m>1$, consistency is not guaranteed by our argument. An inconsistent system excludes this family of shifts for the chosen targets.

For a sufficient test on all smooth invariant divisors, one may instead form $M$ and its diagonal on an integral basis of the invariant bulk divisor lattice. Both depend linearly on the target class $[D]$.

Appendix~\ref{sec:applications} extends the search to other invariant discrete backgrounds, including choices outside the half-sum family.

\subsection{Full worldvolume tests for a shifted background}

After the inherited tests, the remaining autochthonous periods must also be checked. For any such integral class $\gamma$, Eqs.~\eqref{eq:corrected-eta} and \eqref{eq:difference-parity} show that $\int_\gamma\tilde\eta_D^\star$ is an integer precisely when
\begin{equation}
 \boxed{
 \tfrac12\delta_\gamma^2
 -\int_{\gamma+\sigma_*\gamma}\iota^*[A]
 \equiv0\pmod2.
 }
 \label{eq:full-corrected-test}
\end{equation}
The same calculation also identifies an inherited odd flux at the original background:
\begin{equation}
 \eta_D^\star
 +\tfrac12\iota^*\bigl([A]-\sigma^*[A]\bigr)
 =\tilde\eta_D^\star+\iota^*[A].
 \label{eq:bulk-flux-background-relation}
\end{equation}
Since $\iota^*[A]$ is integral, the flux $\mathcal F_D=\tfrac12\iota^*([A]-\sigma^*[A])$ at $B_{2,+}^\star$ satisfies the same period tests as zero flux at $\tilde B_{2,+}^\star$. The general odd-flux construction of Theorem~2 also allows fluxes that are not inherited from the bulk.

\section{Conclusions}
We have shown that the single bulk background $B_{2,+}^{\star}=\tfrac12[R]$ permits Freed--Witten period quantization on any collection of smooth connected divisors and their orientifold images, under the assumptions of Sec.~\ref{sec:quantization}. The worldvolume fluxes are chosen individually, obey the orientifold image relation, and satisfy quantization on the full two-cycle lattices, including inherited and autochthonous classes. No restriction on $h^{1,1}_-(X)$ is required.

The essential ingredient is the invariant-period theorem for setwise invariant divisors. On every such divisor, $\eta_D^\star$ has integer periods on every invariant integral two-cycle class, without requiring an invariant or holomorphic representative. Purely $\sigma$-odd classes have zero period, so any non-integer periods occur only on classes of mixed $\sigma$-parity. This invariant-period statement is precisely the condition needed to complete quantization with an odd flux. For exchanged pairs, a quantized choice on one member also quantizes its orientifold image. The worldvolume fluxes need not be inherited from the bulk. On O7 components, zero worldvolume flux already satisfies quantization on every integral two-cycle.

On a setwise invariant divisor, the harmonic representative of the odd flux is primitive for an invariant K\"ahler metric. When $h^{2,0}_-(D)=0$, it is also of type $(1,1)$ and therefore satisfies the flux conditions for a supersymmetric instanton. This includes every smooth rigid invariant divisor, without requiring $h^{1,1}_-(D)=0$. If $h^{2,0}_-(D)\neq0$, the $(1,1)$ condition must be tested separately. For exchanged divisors, the image relation alone does not ensure primitivity on either member. The Chan--Paton projection and the neutral and charged zero modes remain additional conditions for a contributing instanton.

We have also studied when invariant divisors permit $\mathcal F_D=0$. This requires the bulk Freed--Witten class itself to have integer periods on the full worldvolume lattice. At $B_{2,+}^{\star}$, only classes of mixed $\sigma$-parity require further tests, and the criterion $\delta_\gamma^2\in4\Z$ determines whether each period is integral. The full condition is automatic when the $\sigma$-odd second cohomology of $D$ vanishes.

On inherited lattices of invariant divisors, $B_{2,+}^{\star}$ already works whenever $h^{1,1}_-(X)=0$. Common half-sum shifts extend this inherited guarantee to $h^{1,1}_-(X)\leq1$, and also to higher odd rank when at most one pair is exchanged in an integral bulk basis. These inherited guarantees require no conditions on $h^{2,0}(D)$ or $h^{1,1}_-(D)$. For several exchanged pairs, the binary equations account for the effects of each shift on the other periods and determine which common combinations work. The remaining autochthonous periods must also be integral to retain zero flux on the full worldvolume. The double-octic example makes this distinction explicit: every inherited period is integral at $B_{2,+}^{\star}$, while some autochthonous periods are half-integer. The alternative zero-flux background in that example lies outside the half-sum family.

The common background $B_{2,+}^{\star}$ can therefore be used for instanton quantization while exploring D7 configurations beyond componentwise tadpole cancellation. The exchanged-pair construction also ensures period quantization for smooth D7 divisors that are not individually invariant. This freedom matters for chiral model building, since the componentwise prescription can obstruct a net chiral Standard Model quark-doublet spectrum~\cite{Jain:2026wyl}. Numerical tests and applications to axion spectra generated by rigid ED3 instantons will be presented separately~\cite{Jain2026:to_appear_paper1}.

\acknowledgements
This work is supported by the U.S. Department of Energy (DOE) under Grant No. DE-SC0010103. I am thankful to Federico Carta for useful discussions, and to 
Andreas Schachner for helpful comments on the manuscript.

\appendix

\section{Computing other possible zero-flux backgrounds}
\label{sec:applications}
For setwise invariant target divisors, we now search over all invariant discrete backgrounds with $\mathcal F_D=0$, allowing choices outside the family of shifts \eqref{eq:corrected-B}.\footnote{In Ref.~\cite{Louis:2012nb}, $[R]=9[D_1]+4[D_z]$ and $B_{2,+}=\tfrac12[D_1]$, which equals $B_{2,+}^{\star}-4[D_1]-2[D_z]$ and is therefore an integral shift of our background. In the $h^{1,1}=11$ example of Ref.~\cite{Crino:2022zjk}, Eq.~(5.7) instead chooses $B_{2,+}=B_{2,+}^{\star}-\tfrac12[D_5]$. This different discrete background permits vanishing gauge-invariant flux on five D7 stacks supported on O7 components.} We first impose the inherited period conditions and then add any further worldvolume tests. The calculation applies to any smooth involution and can be used for orientifolds of Kreuzer--Skarke hypersurfaces \cite{Kreuzer:2000xy,Moritz:2023jdb}, including systematic collections of divisor exchanges \cite{Altman:2021pyc,Cao:2024oqx}.

\subsection{The full inherited system}

Set $B_{2,-}=0$. Let $\omega_a$, $a=1,\ldots,n_+$, be an integral basis of the invariant bulk divisor lattice, where $n_+=h^{1,1}_+(X)$. Modulo integral shifts, write
\begin{equation}
 B_{2,+}(x)=\tfrac12x^a\omega_a,
 \qquad x^a\in\Ztwo,
 \label{eq:binary-background}
\end{equation}
and $[D_i]=C_i{}^a\omega_a$ with integer $C_i{}^a$. Then
\begin{equation}
 \eta_{D_i}(x)
 =\tfrac12\bigl(x^a+C_i{}^a\bigr)\iota_i^*\omega_a.
 \label{eq:eta-x}
\end{equation}

Use a full integral bulk basis $[e_A]$, without requiring its elements to have definite $\sigma$-parity. Every inherited period is tested by
\begin{equation}
 Ax=q\quad\text{over }\Ztwo,
 \label{eq:xor-system}
\end{equation}
with rows
\begin{align}
 A^{(i)}_{Aa}&=\int_X[D_i][e_A]\omega_a\bmod2,\nonumber\\
 q^{(i)}_A&=\int_X[D_i]^2[e_A]\bmod2.
 \label{eq:full-inherited-rows}
\end{align}

When $m=0$, the known solution is $x_R=[R]\bmod2$ in the invariant basis. When $m=1$, the one-pair argument supplies the known solution from $[R]-[A_1]-\sigma^*[A_1]$. More generally, Eq.~\eqref{eq:common-system} can provide a common solution for the selected collection. The unrestricted system \eqref{eq:xor-system} may also admit solutions outside that family of shifts.

If $x_0$ is one solution, all other solutions are
\begin{equation}
 x=x_0+v,
 \qquad Av=0\quad\text{over }\Ztwo.
 \label{eq:affine}
\end{equation}
Gaussian elimination both tests consistency and finds these solutions. Row addition is exclusive-OR (XOR): $1+1=0$. If the system is consistent and $A$ has binary rank $s$, there are $2^{n_+-s}$ backgrounds satisfying all tested periods. Appendix~\ref{app:xor-example} gives a worked example.

For $h^{1,1}_{-}(X)=0$, choosing $[e_A]=\omega_A$ recovers
\begin{align}
 A^{(i)}_{ba}&=C_i{}^c k_{cba}\bmod2,\nonumber\\
 q^{(i)}_b&=C_i{}^c C_i{}^d k_{cdb}\bmod2,
 \qquad k_{abc}=\int_X\omega_a\omega_b\omega_c.
 \label{eq:even-basis-rows}
\end{align}
When $h^{1,1}_{-}(X)>0$, the full basis in Eq.~\eqref{eq:full-inherited-rows} is essential.

\subsection{Further periods and physical conditions}

For any integral two-cycle class $\gamma$ on $D_i$, define its integer bulk charges
\begin{equation}
 n_a(\gamma)=\int_\gamma\iota_i^*\omega_a.
\end{equation}
Its period is
\begin{equation}
 \int_\gamma\eta_{D_i}(x)
 =\tfrac12\bigl(x^a+C_i{}^a\bigr)n_a(\gamma).
 \label{eq:other-periods}
\end{equation}
Thus integrality adds the row
\begin{equation}
 n_a(\gamma)x^a=C_i{}^a n_a(\gamma)\pmod2.
 \label{eq:additional-row}
\end{equation}
An integral basis of worldvolume classes tests every period. For exchanged curves, invariance of $\omega_a$ gives $n_a(\sigma(C))=n_a(C)$: either component supplies the same row, while their sum doubles every charge and gives $0=0$ modulo two. A class trivial in $X$ has all $n_a=0$ and imposes no additional condition. An autochthonous class need not be trivial in $X$.

Starting from a known solution, shifts must preserve the rows already satisfied. Substituting $x=x_0+v$ therefore requires $Av=0$ together with the additional equations. If $A$ has rank $n_+$, there is no nontrivial shift within Eq.~\eqref{eq:binary-background}. The enlarged system may have fewer solutions, or none.

An inconsistent enlarged system excludes a common zero-flux background among the candidates being tested. It does not exclude the odd-flux solutions at $B_{2,+}^\star$ guaranteed by Theorem~2. Their supersymmetry condition is discussed in Sec.~\ref{sec:wv-flux}.

\section{An illustrative \texorpdfstring{$h^{1,1}=4$}{h11=4} XOR computation}
\label{app:xor-example}
We illustrate the calculation with four invariant bulk classes, $h^{1,1}=h^{1,1}_+=4$. The following intersection data modulo two are illustrative input, rather than a claim about a specified Calabi--Yau geometry.

For one smooth connected invariant divisor, take
\begin{equation}
 [D]=\omega_1,
 \qquad [R]=\omega_2+\omega_3+\omega_4.
\end{equation}
The configuration $B_{2,+}^{\star}$ corresponds to $x_R=(0,1,1,1)^T$. Suppose
\begin{equation}
 A_{ba}:=\int_X[D]\,\omega_b\omega_a\bmod2
 =\begin{pmatrix}
 1&1&1&1\\
 1&1&0&0\\
 1&0&1&0\\
 1&0&0&1
 \end{pmatrix}.
\end{equation}
Since $[D]=\omega_1$, the right-hand side is the first column:
\begin{equation}
 q_b=A_{b1},
 \qquad q=(1,1,1,1)^T.
\end{equation}
Direct substitution verifies $Ax_R=q$. Gaussian elimination over $\Ztwo$ gives
\begin{equation}
 \left(\begin{array}{cccc|c}
 1&1&1&1&1\\
 1&1&0&0&1\\
 1&0&1&0&1\\
 1&0&0&1&1
 \end{array}\right)
 \longrightarrow
 \left(\begin{array}{cccc|c}
 1&0&0&1&1\\
 0&1&0&1&0\\
 0&0&1&1&0\\
 0&0&0&0&0
 \end{array}\right).
 \label{eq:example-reduction}
\end{equation}
Every row addition is an XOR. For example, adding the original second and third equations cancels $x^1$ and gives $x^2+x^3=0$, since $1+1=0$.

There are three independent equations and one free variable. Setting $x^4=t\in\Ztwo$, the solutions are
\begin{equation}
 x=(1+t,t,t,t)^T.
\end{equation}
Thus there are $2^{4-3}=2$ backgrounds:
\begin{align}
 t=0:\quad B_{2,+}^{(0)}&=\tfrac12\omega_1,\nonumber\\
 t=1:\quad B_{2,+}^{(1)}&=\tfrac12(\omega_2+\omega_3+\omega_4)
 =B_{2,+}^{\star}.
\end{align}
Changing $t$ adds $v=(1,1,1,1)^T$, with $Av=0$. Both choices satisfy the four inherited tests.

To illustrate an additional test, suppose an integral class $\gamma$ has charges $n_a(\gamma)=(0,0,0,1)$. Its period is $x^4/2$, so integrality requires $x^4=0$ and leaves only $B_{2,+}^{(0)}$. Such a class would have mixed $\sigma$-parity: its half-integer period at $B_{2,+}^{\star}$ excludes invariance, and its nonzero period excludes $\sigma$-odd parity.

\section{Choosing a common shift: a two-pair example}
\label{app:lattice-shifts}
We illustrate how two exchanged-pair shifts can affect one another. The binary matrices below are illustrative data. Appendix~\ref{app:exchange} gives a compact Calabi--Yau example of the one-pair construction.

Take two exchanged pairs in a full integral basis, $[e_1]\leftrightarrow[e'_1]$ and $[e_2]\leftrightarrow[e'_2]$, with all remaining basis classes invariant. Choose $[A_1]=[e_1]$ and $[A_2]=[e_2]$. The four candidate backgrounds are
\begin{align}
 \tilde B_{2,+}^{\star}
 =\tfrac12[R]
 &-\tfrac{z^1}{2}([e_1]+[e'_1])\nonumber\\
 &-\tfrac{z^2}{2}([e_2]+[e'_2]),
 \qquad z^1,z^2\in\Ztwo.
\end{align}

Suppose the pairings on two target divisors reduce to
\begin{equation}
 M^{(1)}=\begin{pmatrix}1&0\\0&0\end{pmatrix},
 \qquad
 M^{(2)}=\begin{pmatrix}1&1\\1&1\end{pmatrix}.
\end{equation}
Their diagonal vectors are $d^{(1)}=(1,0)^T$ and $d^{(2)}=(1,1)^T$. The first target requires $z^1=1$. The second requires $z^1+z^2=1$. Stacking and reducing gives
\begin{equation}
 \left(\begin{array}{cc|c}
 1&0&1\\0&0&0\\1&1&1\\1&1&1
 \end{array}\right)
 \longrightarrow
 \left(\begin{array}{cc|c}
 1&0&1\\0&1&0\\0&0&0\\0&0&0
 \end{array}\right).
\end{equation}
The unique common solution is $z=(1,0)^T$, so $[A]=[A_1]=[e_1]$. It selects
\begin{equation}
 \tilde B_{2,+}^{\star}
 =\tfrac12[R]-\tfrac12([e_1]+[e'_1]).
\end{equation}
Including both shifts, $z=(1,1)^T$, would fail the second target.

For comparison, a further block
\begin{equation}
 M^{(3)}=\begin{pmatrix}0&1\\1&0\end{pmatrix},
 \qquad d^{(3)}=(0,0)^T,
\end{equation}
would require $z^1=z^2=0$ and contradict the first target. For any symmetric binary matrix $M$, the system $Mz=\operatorname{diag}(M)$ has at least one solution~\cite{Minevich:2012matrix}. Thus each target admits a solution individually, but the three targets have no common solution. Establishing whether such a pattern occurs in a Calabi--Yau geometry requires geometric intersection data.

\section{Constructing an odd worldvolume flux}
\label{app:odd-flux-construction}
For a setwise invariant divisor, the completion in Proposition~1 can be found by an integer linear system once the full worldvolume lattice is known. Choose an integral basis $\omega_i$ of $H^2(D,\Z)_{\mathrm{free}}$ and write
\begin{equation}
 \sigma^*\omega_j=(T_D)^i{}_j\omega_i,
 \qquad 2\eta_D^\star=k^i\omega_i.
\end{equation}
Both $T_D$ and $k$ have integer entries. Writing $\ell_D=n^i\omega_i$, the condition $\ell_D+\sigma^*\ell_D=2\eta_D^\star$ becomes
\begin{equation}
 (I+T_D)n=k,\qquad n\in\Z^{b_2(D)}.
 \label{eq:integer-flux-construction}
\end{equation}
Here $b_2(D)=\dim_{\mathbb R}H^2(D,\mathbb R)$ is the second Betti number, and $I$ is the $b_2(D)\times b_2(D)$ identity matrix. Theorem~2 guarantees an integer solution. For any such solution,
\begin{equation}
 \mathcal F_D=\sum_i\left(n^i-\tfrac12k^i\right)\omega_i
 =\tfrac12\sum_i\bigl((I-T_D)n\bigr)^i\omega_i.
\end{equation}
This requires the worldvolume matrix $T_D$ and the coefficients of $\eta_D^\star$ in a full integral worldvolume basis. The bulk toric divisor lattice alone need not supply these data.

For a simple period example, take two integral basis cycles $a,b$ exchanged by $\sigma$, with
\begin{equation}
 \int_a\eta=\int_b\eta=\tfrac12.
\end{equation}
The invariant subgroup is generated by $a+b$, whose period is one. Its basis can be extended to the full integral cycle basis $\{a+b,a\}$, since $b=(a+b)-a$. Choose an integral class with periods
\begin{equation}
 \int_a\ell=1,\qquad \int_b\ell=0.
\end{equation}
Its $\sigma$-even projection is $\eta$, while its $\sigma$-odd projection has periods $\tfrac12$ and $-\tfrac12$ on $a,b$. The completed periods are therefore $1$ and $0$. The basis $\{a+b,a-b\}$ would not be a full integral basis, since recovering $a$ or $b$ would require halves.

\section{Zero flux and divisor exchange}
\label{app:exchange}
Consider the compact hypersurface $Q_4^{(dP_9)}$ of Ref.~\cite{Blumenhagen:2008zz}. Its involution has $h^{1,1}_+=4$, $h^{1,1}_-=1$, exchanges $D_7$ and $D_8$, and preserves $D_5,D_6,D_9$. There is one O7 divisor, of class
\begin{equation}
 [R]=[D_5]+[D_7]+[D_8].
\end{equation}
The divisor $D_6$ is a smooth rigid $dP_9$. An $O(1)$ instanton configuration on it is explicitly discussed in Ref.~\cite{Green:2009qk}. This geometry illustrates how the one-pair shift of Theorem~3 removes a half-integer inherited period at $B_{2,+}^{\star}$.

Take the invariant target $D=D_6$ and the inherited curve $C=D_6\cap D_7$. Equations~(28)--(29) of Ref.~\cite{Green:2009qk} give
\begin{align}
 \int_X[D_6]^2[D_7]&=-1,
 &\int_X[D_5][D_6][D_7]&=1,\nonumber\\
 \int_X[D_6][D_7]^2&=-1,
 &\int_X[D_6][D_7][D_8]&=0.
\end{align}
Consequently,
\begin{align}
 I(D_6,D_7)
 &=\tfrac12\int_X[D_6][D_7]\bigl([D_6]+[R]\bigr)\nonumber\\
 &=\tfrac12(-1+1-1+0)=-\tfrac12.
\end{align}
At $B_{2,+}^{\star}=\tfrac12[R]$ with $\mathcal F=0$, this gives
\begin{equation}
 \int_C\eta_{D_6}^{\star}=-\tfrac12.
\end{equation}
The image $C'=D_6\cap D_8$ has the same period, while the invariant sum has period $-1$. This is an inherited failure on a rigid invariant divisor, rather than an additional autochthonous period.

For comparison, the alternative background $B_{2,+}=\tfrac12[D_5]$ used in Ref.~\cite{Green:2009qk} gives
\begin{equation}
 \int_X[D_6][E]\bigl([D_6]+[D_5]\bigr)=0
\end{equation}
for each basis class $[E]=[D_5],\ldots,[D_9]$. All inherited periods on $D_6$ then vanish. This alternative is precisely the one-pair shift:
\[
 B_{2,+}^{\star}
 -\tfrac12\bigl([D_7]+[D_8]\bigr)
 =\tfrac12[D_5].
\]
Theorem~3 therefore guarantees every inherited period on every smooth connected invariant divisor in this geometry.

The shift system reduces to one bit. Choosing $[A_1]=[D_7]$ gives
\begin{equation}
 M^{(D_6)}=(1),\qquad d^{(D_6)}=(1),
\end{equation}
since
\[
 \int_X[D_6][D_7]\bigl([D_7]+[D_8]\bigr)=-1.
\]
The equation is $z^1=1$, which gives $[A]=[D_7]$ directly.

The same curve $C=D_6\cap D_7$, now regarded as a curve on the exchanged divisor $D_7$, gives
\begin{align}
 \int_C\eta_{D_7}^\star
 &=\tfrac12\int_X[D_7][D_6]\bigl([R]+[D_7]\bigr)\nonumber\\
 &=\tfrac12\bigl(1-2+0\bigr)=-\tfrac12.
\end{align}
Thus zero worldvolume flux fails on the exchanged pair $D_7,D_8$ at $B_{2,+}^{\star}$. Nevertheless, the choice $\mathcal F_{D_7}=-\eta_{D_7}^\star$ and $\mathcal F_{D_8}=\eta_{D_8}^\star$ satisfies the image relation and quantizes all periods, as in Theorem~2.

\section{A double-octic example}
\label{app:double-octic}
We use the double-octic construction of Ref.~\cite{Collinucci:2012as} to illustrate a case in which inherited periods permit zero flux but autochthonous periods do not. Consider
\begin{equation}
 X:\quad w^2=x_1^8-x_2^8+x_3^8+x_4^8
 \subset\mathbb P^4_{1,1,1,1,4}.
\end{equation}
This is smooth. The coordinate $w$ has weight four. The projection $\pi:X\to\mathbb P^3$ onto the coordinates $[x_1:x_2:x_3:x_4]$ is a double cover, with involution $\sigma:w\mapsto-w$. If $H$ is the pullback of the hyperplane class, then\footnote{Using $[X]=8H$ in the weighted projective ambient space gives $\int_XH^3=8\int_{\mathbb P^4_{1,1,1,1,4}}H^4=2$.}
\begin{equation}
 \int_XH^3=2,
 \qquad [R]=4H.
\end{equation}
The bulk divisor lattice is generated by the invariant class $H$, so $h^{1,1}_{-}(X)=0$.

Take the smooth connected invariant divisor
\begin{equation}
 D=X\cap\{x_3=x_1-x_2\},
 \qquad h=\iota^*H.
\end{equation}
The image of $D$ is the plane $P=\{x_3=x_1-x_2\}\simeq\mathbb P^2$ in $\mathbb P^3$. The restricted double cover $D\to P$ has equation
\begin{align}
 w^2&=G(x_1,x_2,x_4),\nonumber\\
 G&=x_1^8-x_2^8+(x_1-x_2)^8+x_4^8.
\end{align}
Its branch curve is $G=0$ in $P$, above which the two values of $w$ coincide at $w=0$. The preimage of this curve is $D\cap R$. The surface $D$ itself is not contained in $R$. The branch curve is smooth because the binary form $x_1^8-x_2^8+(x_1-x_2)^8$ has no repeated projective root, so the double-cover surface $D$ is smooth as well. Since $[D]=H$,
\begin{equation}
 c_1(K_D)=h,
 \qquad \int_Dh^2=2.
\end{equation}
At the common background,
\begin{equation}
 B_{2,+}^{\star}=2H,
 \qquad \eta_D^{\star}=\tfrac52h.
 \label{eq:double-octic-eta}
\end{equation}

The two curves
\begin{equation}
 C_\pm:\quad x_1=x_2,\quad x_3=0,\quad w=\pm x_4^4
\end{equation}
lie on $D$ and are exchanged by $\sigma$. Each maps isomorphically to the line $\{x_1=x_2,\ x_3=0\}$ in $\mathbb P^3$. Thus $C_\pm\simeq\mathbb P^1$. A generic hyperplane intersects this line once. Since $h$ is the restricted pullback of the hyperplane class, $\int_{C_\pm}h=1$. Their sum represents the inherited hyperplane intersection class $\gamma_H$, for which $\int_{\gamma_H}h=2$. Therefore
\begin{equation}
 \int_{C_\pm}\eta_D^{\star}=\tfrac52,
 \qquad
 \int_{C_++C_-}\eta_D^{\star}=5.
 \label{eq:double-octic-periods}
\end{equation}
Every inherited class is an integer multiple of $\gamma_H$ and passes, while the individual component periods fail.

Write $\gamma_\pm=[C_\pm]\in H_2(D,\Z)$. Each $C_\pm$ is rational and satisfies $\int_{C_\pm}c_1(K_D)=\int_{C_\pm}h=1$. Adjunction on $D$ therefore gives
\[
 \gamma_\pm^2
 =2g(C_\pm)-2-\int_{C_\pm}c_1(K_D)
 =-2-1=-3.
\]
Since $(\gamma_++\gamma_-)^2=2$, it follows that $\gamma_+\cdot\gamma_-=4$. In particular, the two classes are distinct. Thus their classes, as well as their representatives, are exchanged. Neither component class is an integer multiple of $\gamma_H$. They are autochthonous. For $\gamma=\gamma_+$, the $\sigma$-odd difference has square $-14$, agreeing with Eq.~\eqref{eq:full-criterion}.

Theorem~2 can be realized explicitly on this surface. Let $c_\pm\in H^2(D,\Z)$ be the integral divisor classes of the curves $C_\pm$ on $D$. Their sum is $h$, and the involution exchanges them. The flux
\begin{equation}
 \mathcal F_D=\tfrac12(c_+-c_-)
 \label{eq:double-octic-odd-flux}
\end{equation}
is $\sigma$-odd and of type $(1,1)$. At the original bulk background,
\begin{equation}
 \eta_D^\star+\mathcal F_D
 =3c_++2c_-\in H^2(D,\Z).
 \label{eq:double-octic-completion}
\end{equation}
The two component periods become $-1$ and $6$, respectively. Since $c_\pm$ are integral classes, this identity proves quantization on every integral two-cycle of $D$.

The alternative background $B_{2,+}=\tfrac12H$ gives $\eta_D=h$, with integer periods on every integral worldvolume two-cycle. This background lies outside the family \eqref{eq:corrected-B} in this example. Since $h^{1,1}_{-}(X)=0$, every integral $[A]$ satisfies $\sigma^*[A]=[A]$, and the shift changes $B_{2,+}^{\star}$ only by an integral shift. It cannot repair the half-integer component periods. Thus no $[A]$ in the proposed family satisfies all worldvolume periods with $\mathcal F_D=0$ on this $D$, even though its full inherited lattice passes. 

\section{Invariant autochthonous classes}
\label{app:invariant-autochthonous}
Invariant worldvolume classes need not be inherited from bulk divisors. A direct example is provided by the Fermat quintic
\begin{equation}
 X:\quad x_0^5+x_1^5+x_2^5+x_3^5+x_4^5=0
 \subset\mathbb P^4,
\end{equation}
with involution
\begin{equation}
 \sigma:[x_0:x_1:x_2:x_3:x_4]
 \longmapsto[x_1:x_0:x_2:x_3:x_4].
\end{equation}
This preserves the quintic and sends $\Omega_X$ to $-\Omega_X$. Its O7 divisor is $R=X\cap\{x_0=x_1\}$, and it has the isolated fixed point $[1:-1:0:0:0]$. Both $X$ and the invariant divisor $D=X\cap\{x_4=0\}$ are smooth. The divisor $D$ is not fixed pointwise.

Let $H$ denote the hyperplane class of $X$ and $h=\iota^*H$. The bulk divisor lattice is $\Z H$, so the inherited lattice of $D$ is generated by $\gamma_H$, with
\begin{equation}
 \int_{\gamma_H}h=\int_Dh^2=5.
\end{equation}
Now consider the line
\begin{equation}
 C:\quad x_0+x_1=x_2+x_3=x_4=0.
\end{equation}
It lies on $D$ and is preserved setwise by $\sigma$. Its class is therefore invariant. Since $C$ is a line, $\int_C h=1$, so $[C]$ cannot be an integer multiple of $\gamma_H$. It is autochthonous.

Here $[R]=[D]=H$, so
\begin{equation}
 B_{2,+}^{\star}=\tfrac12H,
 \qquad \eta_D^{\star}=h,
 \qquad \int_C\eta_D^{\star}=1.
\end{equation}
Thus $C$ supplies an explicit invariant autochthonous class covered by Theorem~1 on a divisor outside the O7 locus.

A rigid example appears in Sec.~4.3 of Ref.~\cite{Blumenhagen:2008zz}. The threefold has $h^{1,1}(X)=4$, while the setwise invariant rational elliptic surface $D_5\simeq dP_9$ has $h^{1,1}_+(D_5)=6$. The inherited lattice has rank at most four, so at least two independent invariant classes lie outside its rational span.

\bibliography{bibfile}

\end{document}